\documentclass[aps,reprint,twocolumn,amsmath,amssymb,groupedaddress,prb]{revtex4-1}
\usepackage{graphicx}
\usepackage{dcolumn}
\usepackage{bm}
\usepackage{amsfonts}
\usepackage{xcolor}
\usepackage{multirow}
\setcitestyle{numbers,square}
\usepackage{enumerate}
\usepackage[normalem]{ulem}
\usepackage{color}
\usepackage{hyperref}
\hypersetup{
    colorlinks,%
    citecolor=blue,%
    linkcolor=blue,%
    urlcolor=blue
}

\newcommand{\Fig} [1]  {Fig.~\ref{#1}}

\newcommand{\Tbl} [1]  {Table~\ref{#1}}       

\newcommand{\dgr}   {$^\circ$}

\begin{document}

\title{TCNQ self-assembly driven by molecular coverage over borophene monolayers}

\author{G. H. Silvestre$^{1,2}$ and R. H. Miwa$^{1}$}

 \affiliation{$^{1}$Instituto de F\'isica, Universidade Federal de Uberl\^andia, 38400-902, Uberl\^andia, MG, Brazil}
 \affiliation{$^{2}$ Departamento de F\'isica, Universidade Federal de Ouro Preto, 35400-000, Ouro Preto, MG, Brazil}

\date{\today}

\begin{abstract}

Boron monolayers, also known as borophene, have recently attracted interest due to their electronic properties, e.g. the facility to form various allotropes with interesting properties. In this work, we investigate the adsorption process of the tetracyanoquinodimethane (TCNQ) on the borophene $\beta_{12}$ and $\chi_3$ using the density functional theory (DFT). We observed that molecules bond to the borophene layer through the van der Waals interaction, where, at the low coverage limit, the binding strength of TCNQ / borophene is comparable to that of TCNQ / WSe$_2$. By increasing the molecular coverage, $10^{13}$\,$\rightarrow$\,$10^{14}$\,molecules/cm$^{2}$, we found the (exothermic) formation of self-assembled (SA) structures of TCNQ on borophene, where the molecule-molecule interactions rule the SA process. The structural stability of the SA-TCNQ molecule on borophene was verified via ab initio molecular dynamics simulations. Finally, we show that the formation of the vdW interface leads to the tunability of the hole-doping of the borophene layer by an external electric field. We believe that our results bring an important contribution to the atomic-scale understanding of a powerful electron acceptor molecule, TCNQ, adsorbed on a promising 2D material, borophene.

\end{abstract}

\maketitle

\section{Introduction}

Two-dimensional (2D) materials have been the object of several studies due to their prominent electronic and structural properties. The application of these materials is mainly related to the development of nanoelectronic devices, in addition to applications in phenomena such as catalysis, substrate, and others. Since the discovery of graphene in 2004 \cite{geimNatMat2007}, several other 2D monoatomic materials, such as silicene \cite{tao2015silicene,lew2016silicene}, phosphorene \cite{carvalho2016phosphorene} and borophene \cite{penevNanoLett2012,wuACSNano2012} have been intensively studied. 

Borophene is a 2D sheet of boron atoms displayed in different configurations. In 2015, Mannix et al. successfully synthesized corrugated boron monolayers through the deposition of boron from a source with high purity on an inert silver substrate \cite{mannixScience2015}. A year later, Feng et al. also obtained boron monolayers from the direct evaporation of a pure boron source, giving rise to two new borophene allotropes, both planar with vacancies \cite{fengNatChem2016}. Recently, Ma et al. synthesized borophene bilayers that are more stable than the monolayers \cite{ma2022prediction}. The complexity of the boron allotropes can be attributed to its trivalent electronic configuration \cite{wuACSNano2012}, which gives rise to several electronic properties \cite{jiangNanoEnergy2016,silvestre2019electronic,arabha2020engineered} and borophene's metallic character makes it a potential complement to graphene, hBN and TMDs \cite{tangPRL2007,liuAngWandChem2013,pengJMatChem2016}.

The incorporation of organic molecules in 2D materials has been intensively studied in the last few years. Gobbi et al. emphasise the potential of hybrid organic/inorganic van der Waals heterostructures, which allow for the alteration of 2D material characteristics and the interface of continuous molecular layers with inorganic 2D materials to enable innovative device topologies \cite{gobbi20182d}. Another potential application of this building is organic 2D materials' uses in optoelectronic devices, highlighting their benefits in terms of processing simplicity and molecular diversity \cite{yang20182d}, besides the advantages of mixing 2D and organic molecular materials, which make it possible to design and synthesise 2D organic materials on a large scale with superior optical and electrical properties \cite{yan20232d}. Molecule-based functionalization is also a potent method to achieve the necessary tuning of these novel materials' optical and electronic properties, which is required for many practical applications \cite{daukiya2019chemical,de2015organic,yang2017many,riedl2010structural,li2021self,slassi2023non,bertolazzi2018molecular}.

The self-assembly process is primordial to control the geometric structure and performance of devices \cite{van2016modeling}, and it can lead to the formation of 1D rows and 2D layers with tailor-made properties and functionality \cite{kuhnle2009self}. Overall, self-assembly plays a crucial role in the bottom-up formation of functional nanostructures with desired properties and applications in various fields. This process is determined by factors such as the properties of the building blocks, molecule-molecule interactions, molecule-substrate interactions, and initial spatial configuration \cite{roos2011intermolecular,lombardo2020self}. Among the organic molecules, the 7,7,8,8-tetracyanoquinodymethane (TCNQ) is well-known in the literature as an excellent electron-acceptor material and has also attracted considerable attention as a mixed-valent partner in organic charge-transfer complexes \cite{goetz2014charge}. Its uses are varied and range from optoelectronics \cite{jing2014tuning}, energy storage \cite{wang2023charge}, transport properties \cite{precht2015electronic,hanyu2012rechargeable},and  catalysis \cite{mohammadtaheri2016emerging}.


In this study, we present a comprehensive discussion of the adsorption of TCNQ (C$_{12}$N$_4$H$_4$) molecule in borophene monolayers structures $\beta_{12}$ and the $\chi_{3}$, here labeled as S1 and S2 (TCNQ/S1 and /S2). The occurrence of the self-assembled (SA) TCNQ structures on borophene (SA-TCNQ/borophene) was examined in light of total energy calculations and {\it ab-initio} molecular dynamics (AIMD) simulations. We found that the latter system can be characterized as a 2D vdW heterostructure composed of an SA-TCNQ sheet bonded to a borophene layer.  We show that the TCNQ/ and SA-TCNQ/borophene systems lead to the hole-doping of borophene, where the amount of doping ($\Delta\rho$) can be tuned by an external electric field (EEF).

\section{Computational Details}

All calculations were performed within the density functional theory (DFT), where the exchange-correlation term was described by the generalized gradient approximation (GGA-PBE) \cite{PBE} proposed by Perdew, Burke and Ernzerhof. The equilibrium geometries, total energies, and {\itshape ab initio} molecular dynamics (AIMD) were calculated using the software Vienna Ab-initio Simulation Package (VASP) \cite{vasp1,vasp2}. We have used the super-cell approach, and the image interaction was avoided using a 15 \AA~vacuum perpendicular to the borophene sheet, alongside a dipole correction. The Kohn-Sham orbitals were expanded in a plane-wave basis set with an energy cutoff of 400 eV. The 2D Brillouin Zone is sampled following the Monkhorst-Pack scheme \cite{mp}, using a 6$\times$6$\times$1 gamma-centered mesh. The electron-ion interaction was considered using the Projector Augmented Wave (PAW) method \cite{paw}. In Borophene/TCNQ systems, all atoms were fully relaxed until convergence criteria of 25 meV\AA$^{-1}$~ were achieved meanwhile the cell volumes weren't allowed to relax, and for layered TCNQ, the convergence criteria were the same as previous, but the cell volumes were relaxed until reached pressure smaller than 0.5 kbar. To provide a better description of stacking, we have considered a non-local vdW-DF2\cite{lee2010higher} vdW approach. The Born-Oppenheimer AIMD applied to analyse the dynamic properties of the molecules over the borophene was performed by using the canonical ensemble (NVT) with a Nose-Hoover thermostat, equilibrated at 300, with a simulation time of 10 ps (1 fs time step). 
Post-processing features were obtained using software VESTA \cite{momma2011vesta}, Vaspkit \cite{wang2021vaspkit}, and VASProcar \cite{vasprocar}.

\section{Results and Discussions}

  \begin{figure}
    \includegraphics[width=\columnwidth]{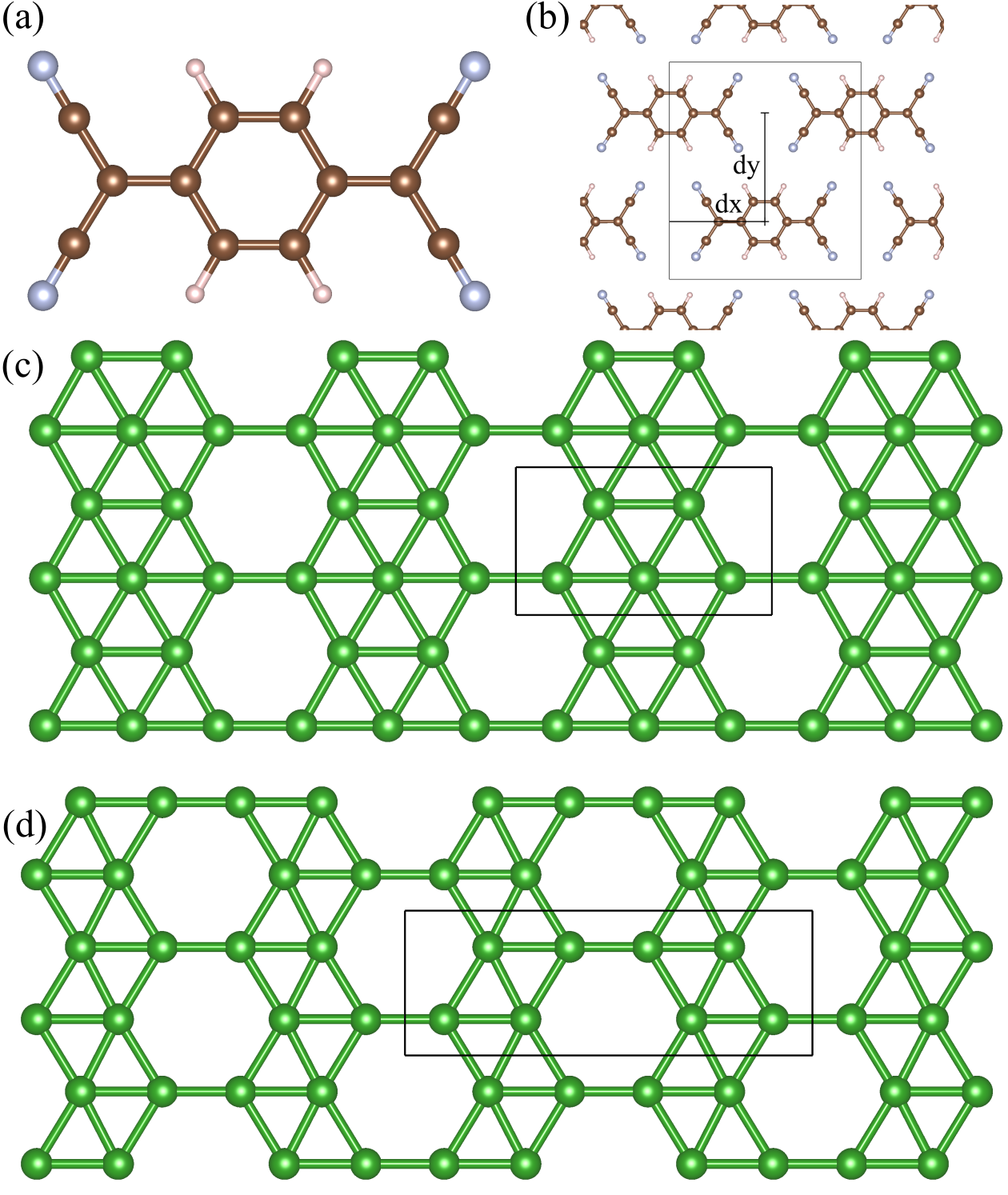}
    \caption{\label{fig1} Structural models of TCNQ (a), S1 (c) and S2 (d) borophene. The hydrogen, boron, carbon and nitrogen are represented by pink, green, brown and grey balls. The unit cell is indicated by the dashed black line in (c) and (d). In (b), the separation distances d$_x$ and d$_y$ between two TCNQ molecules are measured from the benzene ring centre.}
  \end{figure}

  \begin{figure*}
    \includegraphics[width=2\columnwidth]{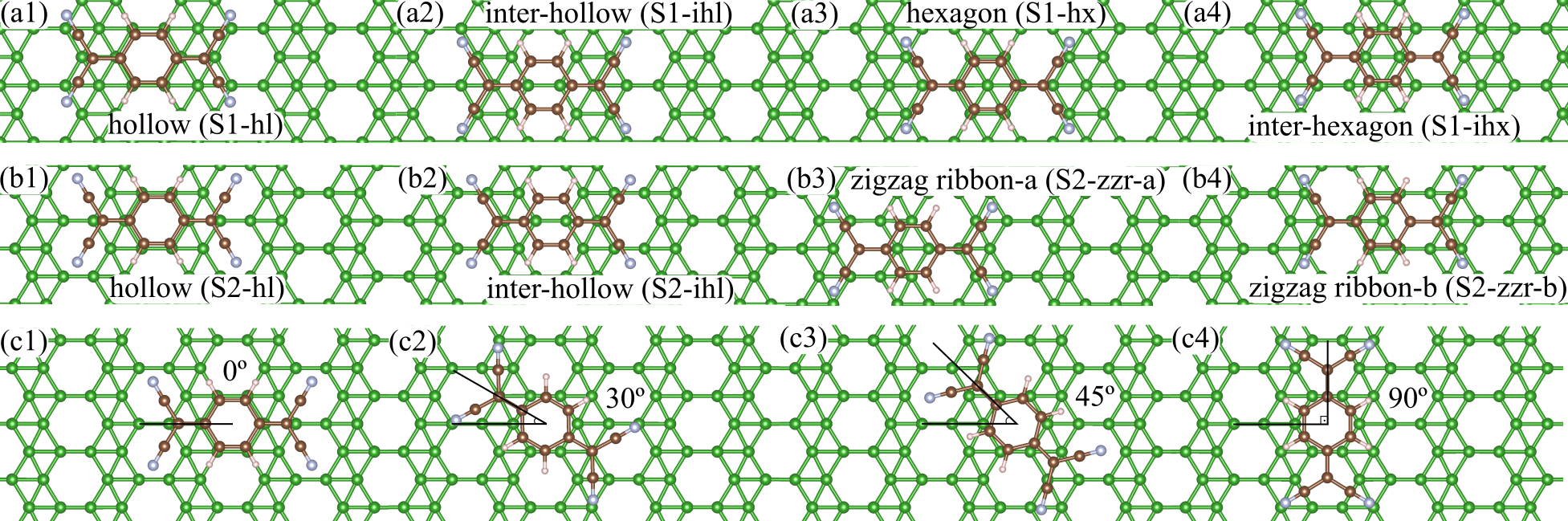}
    \caption{\label{fig2} Adsorption sites of TCNQ in S1: (a1) hollow (S1-hl), (a2) inter-hollow (S1-ihl), (a3) hexagon (S1-hx) and (a4) inter-hexagon (S1-ihx). Adsorption sites of TCNQ in S2: (b1) hollow (S2-hl), (b2) inter-hollow (S2-ihl), (b3) zigzag ribbon-a (S2-zzr-a) and (b4) zigzag ribbon-b (S2-zzr-b). Rotation angles of (c1) 0\dgr~, (c2) 30\dgr~, (c3) 45\dgr~ and (c4) 90\dgr.}
  \end{figure*}

In Fig.\,\ref{fig1}(a) we present the structural model the TCNQ molecule, and in Fig.\,\ref{fig1}(c) and (d) we show the geometries of the planar single layer borophene\,\cite{fengNatChem2016, shuklaPCCP2018}, labeled as S1 and S2, respectively. Although somewhat structurally similar, S1 and S2 present different hole concentrations ($\eta$), namely 1/6 and 1/5. At the equilibrium geometry, our results for the borophene lattice parameters are in good agreement with the previous experimental and theoretical findings \cite{silvestre2019electronic,boroun2019separated}, namely $a =$ 5.07 (8.30)\,\AA~and $b =$ 2.92 (2.95) \AA~for S1 (S2), and both phases are metallic as depicted in \Fig{fig:sm1} of the SM.


\subsection{Energetic and Structural Properties}

Let us start our study by considering the limit of the low coverage of TCNQ molecules adsorbed on the borophene S1 and S2 phases (TCNQ/S1 and TCNQ/S2). To minimize the molecule-molecule interactions, we built a 4$\times$6$\times$1 S1 supercell, corresponding to a TCNQ coverage area of 2.81$\times$10$^{13}$ molecules/cm$^2$; and 3$\times$6$\times$1 S2 supercell, resulting in  2.27$\times$10$^{13}$ molecules/cm$^2$,  guaranteeing at least 10 \AA~between the image of the molecules. For each one of the phases, TCNQ/S1 and /S2, we tested several different adsorption sites by translating and rotating the molecules over the borophene. Some configurations are indicated by \Fig{fig2}. The energetic stability was examined through the calculation of the adsorption energy (E$^{A}$), given by
\begin{equation}
    E^{A} = \frac{1}{n}(E_{sis} - E_{B} - nE_{mol}),
\end{equation}
where $E_{sis}$ is the energy of the borophene-molecule system, $E_B$ and $E_{mol}$ are the energy of the isolated borophene and molecule, respectively, and $n$ is the number of molecules. The results of $E^{A}$ calculated in each site are summarized in \Tbl{bind-1-1}. At the equilibrium geometry, the TCNQ/borophene distance is about 3.2\,\AA, close to the one obtained in the TCNQ/graphene system.


\begin{table}[h!]
\caption{\label{bind-1-1} Adsorption energies ($E^{A}$) of the TCNQ isolated molecule in pristine borophene S1 and S2 phases  (eV/molecule)}
\begin{ruledtabular}
\begin{tabular}{lcccc}
site & \multicolumn{4}{c}{S1} \\
    & 0$^{\circ}$    & 30$^{\circ}$    &    45$^{\circ}$    &  90$^{\circ}$ \\   
  \cline{2-5} \\
  S1-hl & -1.20  & -1.21  & -1.22  & -1.25    \\ 
 S1-ihl & -1.16  & -1.28  & -1.23  & -1.17    \\ 
  S1-hx & -1.29  & -1.23  & -1.26  & -1.20    \\ 
 S1-ihx & -1.31  & -1.27  & -1.23  & -1.20    \\ 
 & \multicolumn{4}{c}{S2} \\
    & 0$^{\circ}$    & 30$^{\circ}$    &    45$^{\circ}$    &  90$^{\circ}$ \\   
  \cline{2-5} \\
  S2-hx & -1.26 & -1.24  & -1.28  & -1.25   \\ 
 S2-ihx & -1.25 & -1.26  & -1.25  & -1.29   \\ 
 S2-zzr-a & -1.25 & -1.33  & -1.30  & -1.32   \\ 
 S2-zzr-b & -1.24 & -1.30  & -1.26  & -1.31   \\ 
\end{tabular}
\end{ruledtabular}
\end{table}

The $E^A$ results show that there is a slight variation of the $E^A$ for TCNQ in the S1 phase, varying from $-1.16$\,eV in ihl-S1 to $-1.31$\,eV in ihx-S1, both in 0$^{\circ}$, resulting in an averaged adsorption energy, $\langle E^A\rangle\pm\Delta\langle E^A\rangle$ of $-1.23\pm 0.04$\,eV. In S2,  we found slightly higher values of adsorption energies, $\langle E^A\rangle\pm\Delta\langle E^A\rangle$\,=\,$-1.27\pm\,0.02$\,eV. These values of $E^A$ are comparable (larger) than that predicted in TCNQ/WSe$_2$ (TCNQ/MoS$_2$) and lower than TCNQ/graphene.  Moreover, it is worth noting that the somewhat similar values of adsorption energy, $\Delta\langle E^A \rangle$ between 0.02 and 0.03\,eV, suggest that the TCNQ molecules can easily diffuse and rotate on the borophene S1 and S2 surfaces, which, similarly to what has been found in TCNQ/graphene\,\cite{de2015organic}, may lead to the formation of self-assembled (SA) structures of TCNQ on the borophene surface.

The formation of molecular self-assemblies at surfaces is driven by a synergistic combination of molecule-surface and molecule-molecule interactions\,\cite{macleod2014molecular}. In this sense, our results of $\Delta\langle E^A\rangle$ above suggest that the molecule-molecule interaction may bring an important contribution to the (possible) formation of SA structures at borophene. Here, the strength of the TCNQ-TCNQ interaction was examined by using a planar free-standing layer of TCNQ, as shown in Fig.\,\ref{fig1}(b), where the total energy was minimized as a function of the lateral coordinates d$_\text{x}$ and d$_\text{y}$. At the equilibrium geometry we found d$_\text{x}$\,=\,6.74\,\AA\, and d$_\text{y}$\,=\,5.94\,\AA, respectively, and molecule-molecule binding energy ($E^B$) of $-0.6$\,eV/molecule\,\footnote{The binding energy per molecule is defined by the difference between the molecule layer energy and the isolated molecule energy $n$-times divided by the number of $n$ molecules, $E^B = \frac{1}{n}(E_{layer}-nE_{TCNQ})$.}. These results are in line with those obtained in Ref.\,\cite{de2015organic}. Then, to minimize the lattice mismatch between the TCNQ layer and borophene S1 and S2, we construct borophene supercells with surface periodicities of 4\,$\times$\,8 (S1) and 4\,$\times$\,5 (S2), which result in a mismatch of 1.5\% (d$_\text{x}$) and 0.3\,\% (d$_\text{y}$) and 0.6\% (d$_\text{x}$) and 2.6\% (d$_\text{y}$), respectively, for a 1\,$\times$\,3 TCNQ layer. Here, lattice constants of the borophene layers have been conserved, while the TCNQ layer was compressed to preserve the periodic boundary condition, leading to the reduction of $E^B$ by about 10\,\%. We obtained adsorption energies of $-1.58$ and $-1.60$\,eV/molecule in S1 and S2, thus revealing that the formation of SA TCNQ structures on the borophene surfaces (SA-TCNQ/S1 and /S2) is an exothermic process. This shows that SA structures of TCNQ molecules on borophene are possible in terms of energy.

  \begin{figure}
    \includegraphics[width=\columnwidth]{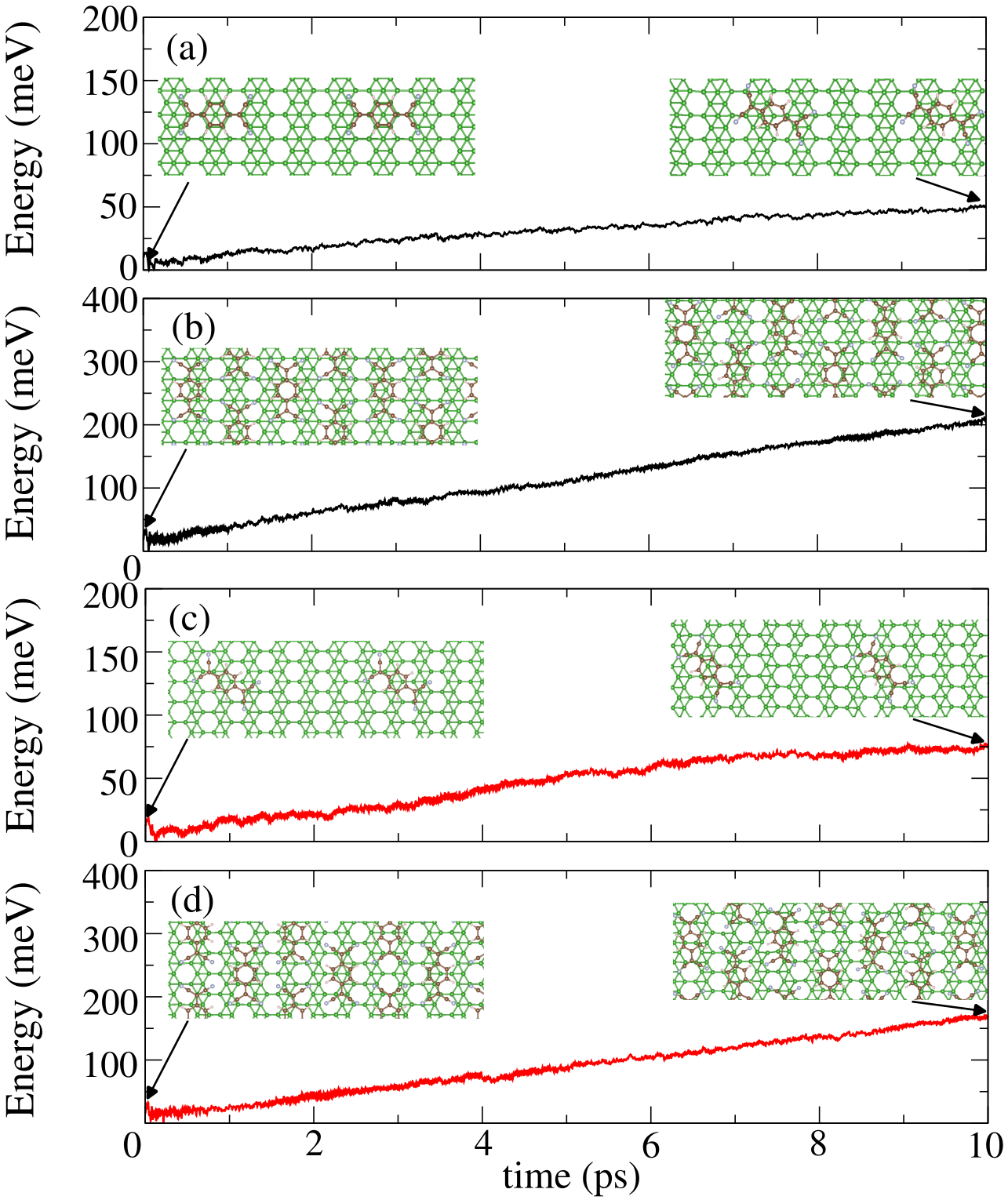}
    \caption{\label{fig3} Ab-initio molecular dynamics (AIMD) of  TCNQ/S1 (a), SA-TCNQ/S1 (b), TCNQ/S2 (c) and SA-TCNQ/S2 (d). In the inset, we presented a snapshot of the initial and final geometry.}
  \end{figure}

Next, to verify the structural stability of the SA-TCNQ/borophene structures, we performed a set of AIMD simulations of SA-TCNQ/S1 and /S2 at an average temperature of 300\,K  during the 10\,ps. In Fig.\,\ref{fig3}, we present the evolution of the SA-TCNQ/borophene systems and (inset) the respective initial and final configurations. Our AIMD results demonstrate that the SA-TCNQ/borophene system forms a structurally stable 2D vdW heterostructure (further details are provided in SM). 


\subsection{Electronic Properties}

\begin{figure}[h!] 
    \includegraphics[width=\columnwidth]{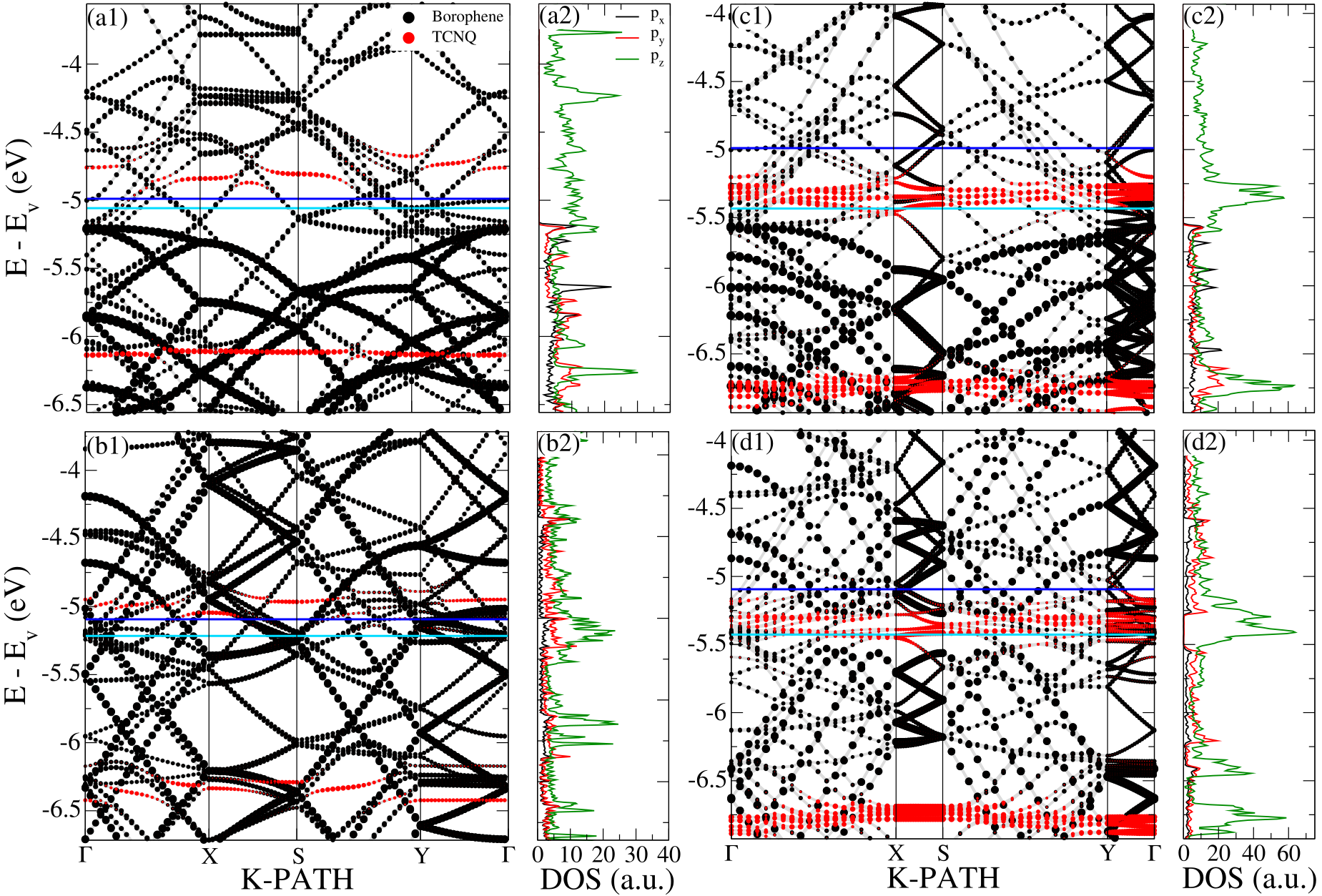}
    \caption{\label{fig4} Projected band structure/PDOS of the isolated TCNQ/S1 (a1/a2) and /S2 (b1/b2) borophene; and the SA-TCNQ/S1 (c1/c2) and /S2 (d1/d2) borophene close to the Fermi level (E$_F \pm$ 1.5 eV). The vacuum level was set to zero; the light blue line indicates the Fermi level, and the dark blue line indicates the Fermi level of the pristine borophene.}
  \end{figure}

TCNQ molecules on 2D platforms may act as an electron acceptor, leading to a hole-doping of the host, like in the well-known TCNQ/graphene, /MoS$_2$, and /Black-phosphorous\,\cite{zhang2016surface}. The orbital projected electronic band structures of TCNQ/S1 and /S2 are shown in Fig.\,\ref{fig4}(a) and (b). It is noteworthy that the metallic band structures of borophene S1 and S2 are maintained. However, in contrast to the pristine S1 and S2 borophene, the final system, TCNQ/borophene, exhibits a downshift of the Fermi level, which increases the work function ($\Phi$). We found $\Phi$\,=\,4.99\,$\rightarrow$\,5.06\,eV (TCNQ/S1), and $\Phi$\,=\,5.10\,$\rightarrow$\,5.22\,eV (TCNQ/S2) as a consequence of the hole doping of borophene by  0.55\,$e$/molecule (1.63$\times$10$^{13} e$/cm$^2$) and 0.58\,$e$/molecule (1.37$\times$10$^{13} e$/cm$^2$), respectively. The amount of charge transferred is comparable with graphene and TMDs, $\approx$ 21\% less than graphene, $\approx$ 5\% greater than WSe$_2$ and 36\% greater than MoS$_2$ \cite{de2015organic,slassi2023non}.

The energy positions of the molecular states (shown by red-filled circles) in TCNQ/borophene and SA-TCNQ/borophene are somewhat similar, as can be seen in Figs.\,\ref{fig4}(a)-(c) and (b)-(d). However, the hole-doping of borophene increases to 1.46\,$e$/molecule (3.08$\times$10$^{13} e$/cm$^2$) and 1.66\,$e$/molecule (3.38$\times$10$^{13} e$/cm$^2$) in SA-TCNQ/S1 and /S2, respectively, which is a consequence of the increase in the amount of adsorbed molecules in the SA systems. In this case, the work function increases to $\Phi$\,= 5.43 eV for both SA-TCNQ/S1 and /S2. Moreover, the dispersionless nature of the molecular states in Figs.\,\ref{fig4}(c) and (b) are noteworthy, allowing us to infer that, despite the TCNQ molecules' lateral proximity in the SA structure, there are no intermolecular wave function overlaps.

  \begin{figure}[h!]
    \includegraphics[width=\columnwidth]{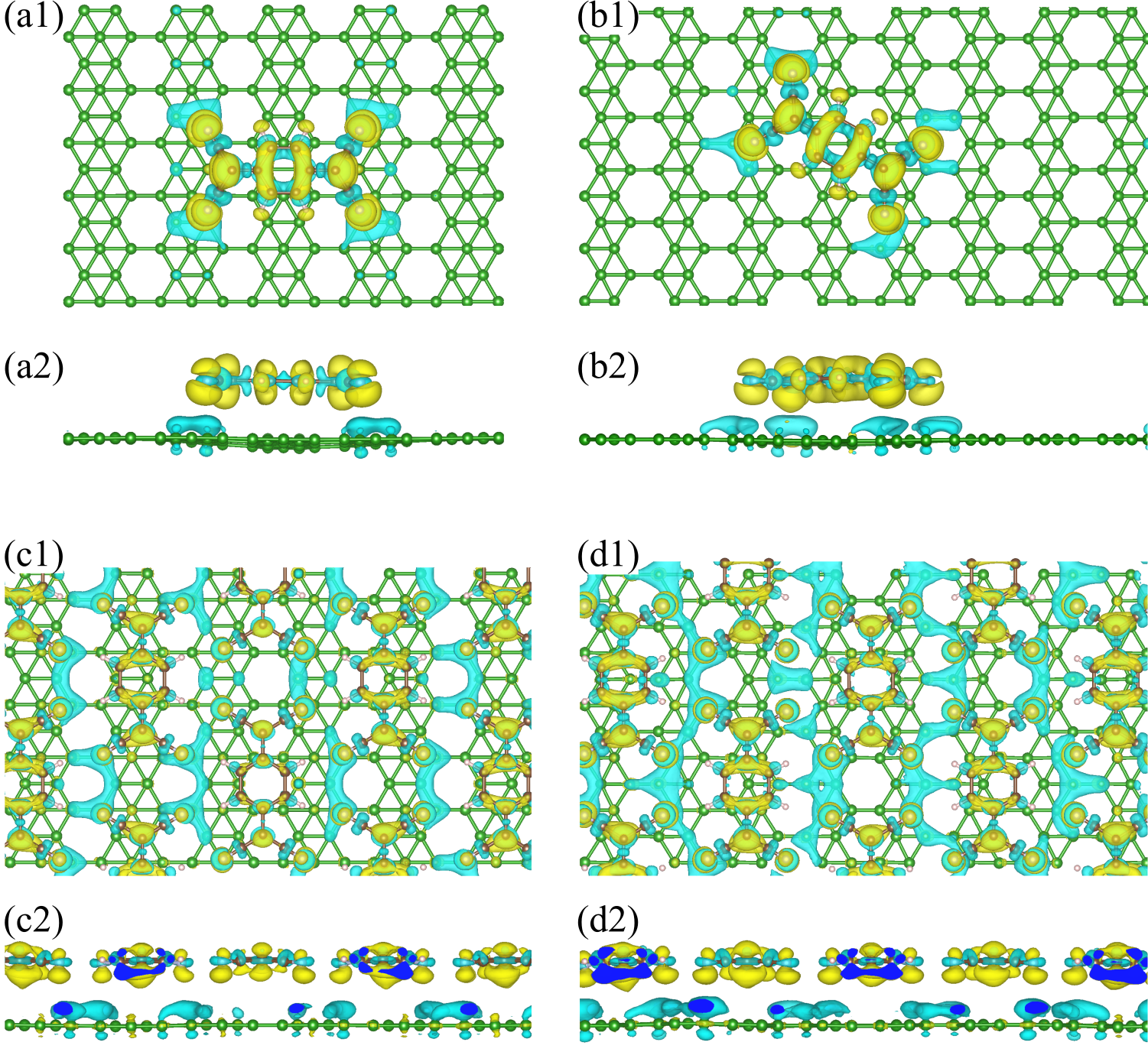}
    \caption{\label{fig5} Charge transfer between the isolated TCNQ/S1 (a) and /S2 (b) and SA-TCNQ/S1 (c) and /S2 (d). The blue region indicates a loss of electrons; meanwhile, the yellow region indicates a gain of electrons. Isosurfaces of 0.5$\times$10$^{-3}$  e/\AA$^3$.}
  \end{figure}

The spatial distribution of the TNCQ-borophene net charge transfers ($\Delta\rho$) is depicted in Fig.\,\ref{fig5}. In both TCNQ/S1 and /S2, the hole-doping of borophene is localized near the molecular adsorption site [Fig.\,\ref{fig5}(a) and (b)], resulting in hole puddles on the borophene surface, which may bring deleterious effects on the electronic transport properties along the borophene surface. On the other hand, the distribution of the hole-doped regions becomes more uniform following the SA molecular arrangement of the TCNQ molecules in SA-TCNQ/borophene, as shown in Fig.\,\ref{fig5}(c) and (d).

  \begin{figure}[h!]
    \includegraphics[width=\columnwidth]{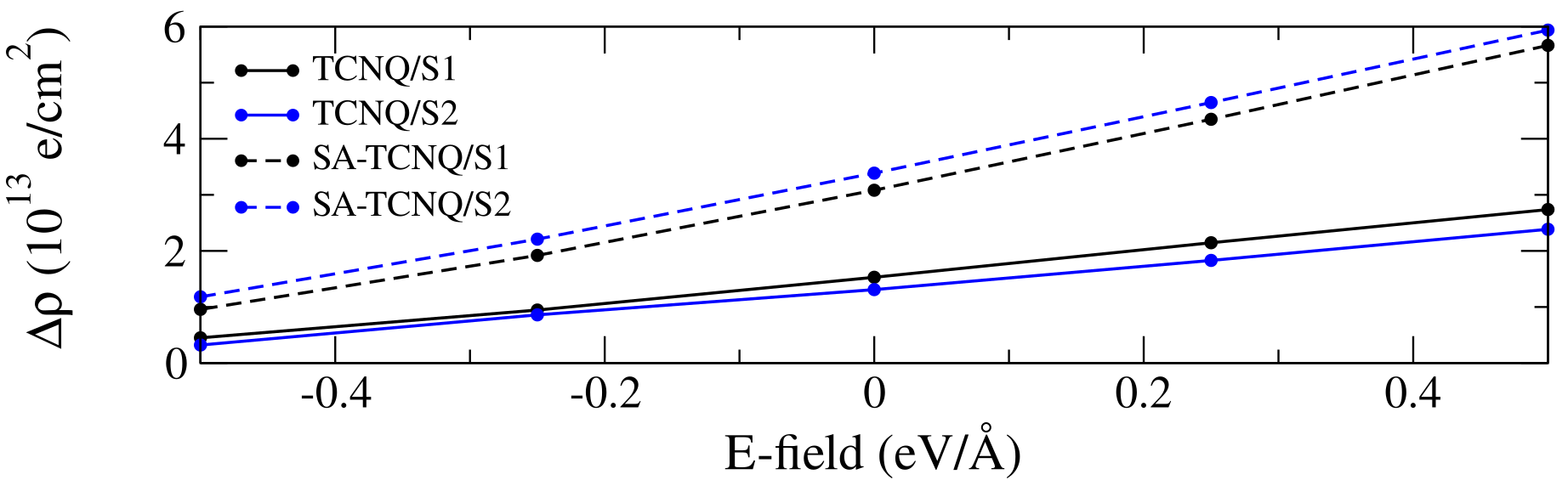}
    \caption{\label{fig6} Amount of doping ($\Delta\rho$) per area of the isolated TCNQ and SA-TCNQ over S1 and S2 borophene.}
  \end{figure}

External agents such as electric fields and mechanical strain are powerful tools to control the electronic properties of materials \cite{souza2022magnetic}. Here, we will focus on the tuning of the net charge transfer process at the TCNQ/ and SA-TCNQ/borophene interfaces by an external electric field (EEF). As presented in Fig.\,\ref{fig6} for an EEF interval of $-0.5$ and $+0.5$ eV/\AA, $\Delta\rho(\text{EEF})$ presents (i) a linear dependence, practically independent of the borophene phase, S1 or S2, where (ii) the SA-TNCQ/borophene structure presents a larger dependence with the applied EEF, which can be attributed to the larger density of molecular states near the Fermi level as shown in Fig.\,\ref{fig4}(c) and (d). In addition, in (i), we can infer that the hole-doping of borophene can be tuned and eventually suppressed upon higher values of negative EEF. 

\section{Summary and Conclusions}

Based on first-principles DFT calculations, we performed a theoretical study of molecular adsorption on borophene. We have considered TCNQ molecules on (experimentally synthesized) two structural phases of borophene. We found that the TCNQ molecules bind to the borophene host through vdW interaction. By examining a large number of TCNQ/borophene configurations, we found that the molecule can be easily diffused over the borophene surface. Further investigation revealed that the (exothermic) formation of self-assembled arrays of TCNQ molecules on borophene is dictated by molecule-molecule interactions. The electronic band structures of the borophene hosts are mostly preserved, where the borophene host becomes hole-doped. Finally, we examined the effect of an external electric field normal to the TCNQ/borophene interface. We verified that the hole-doping of borophene can be tuned and eventually suppressed.

\begin{acknowledgments}

The authors acknowledge financial support from the Brazilian agencies Brazilian Nanocarbon Institute of Science and Technology - INCT Nanocarbono, Coordenação de Aperfeiçoamento de Pessoal de Ensino Superior (CAPES), Conselho Nacional de Desenvolvimento Científico e Tecnológico (CNPq), Fundação de Apoio à Pesquisa de Minas Gerais (FAPEMIG), and Rede Mineira de Pesquisa em Materiais Bidimensionais research projects. G.H.S. acknowledges Universidade Federal de Ouro Preto, FIMAT and Grupo Nano. The authors acknowledge the  Laboratório Nacional de Computação Científica (LNCC-SCAFMat2), Centro Nacional de Processamento de Alto Desempenho (CENAPAD-SP) for computer time.

\end{acknowledgments}

\bibliography{RHMiwa-cellu}


 \section{SUPPLEMENTARY MATERIAL}

0\dgr S1-ihl MD full videos available in: 
\begin{itemize}
    \item \url{https://youtu.be/4nQApcOEoe0}
    \item \url{https://youtu.be/vk0XdpJCkoI}
\end{itemize}

30\dgr S2-zzr-a MD full video available in: 
\begin{itemize}
    \item \url{https://youtu.be/ibxM_nAKkIk}
    \item \url{https://youtu.be/FF_QMKDFc04}
\end{itemize}

90\dgr SA-TCNQ/S1 MD full video available in: 
\begin{itemize}
    \item \url{https://youtu.be/osfNjoY443E}
    \item \url{https://youtu.be/4XbIuUezmlM}
\end{itemize}

90\dgr SA-TCNQ/S2 MD full video available in: 
\begin{itemize}
    \item \url{https://youtu.be/wnoYj3Dy6CU}
    \item \url{https://youtu.be/0DUamJ9wQPA}
\end{itemize}

\begin{figure}[h!]
    \centering
    \includegraphics[width=\columnwidth]{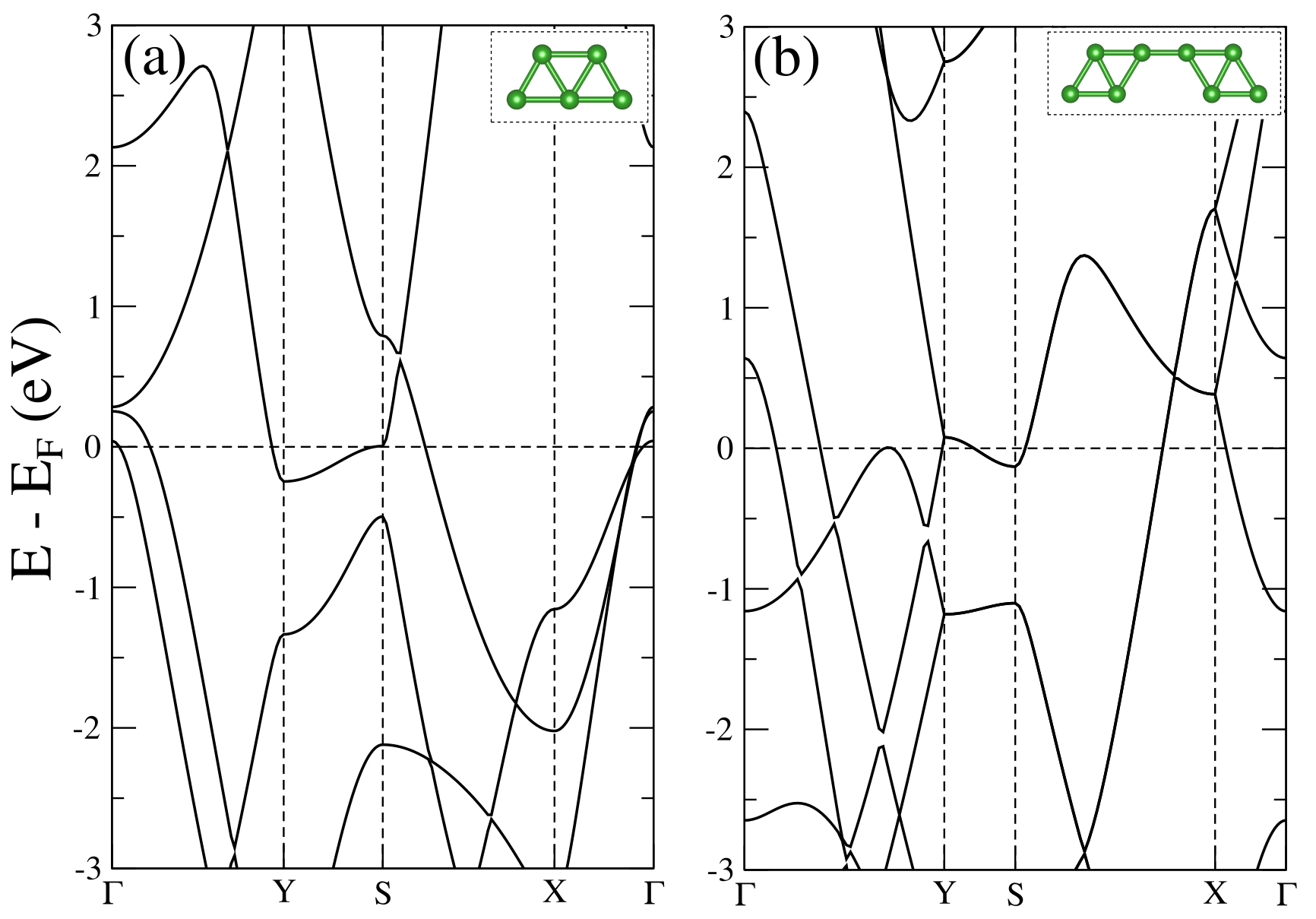}
    \caption{Electronic band structure of borophene S1 (a) and S2 (b).}
    \label{fig:sm1}
\end{figure}

\end{document}